\documentclass[sigconf, authorversion, nonacm]{acmart}

\begin{document}

\title{Long-Term Productivity Based on Science, not Preference}

\author{Spencer Smith}
\orcid{0000-0002-0760-0987}
\affiliation{
  \institution{McMaster University}
  \city{Hamilton}
  \country{Canada}
  }
\email{smiths@mcmaster.ca}

\author{Jacques Carette}
\orcid{0000-0001-8993-9804}
\affiliation{
    \institution{McMaster University}
    \city{Hamilton}
    \country{Canada}}
  \email{carette@mcmaster.ca}

\thanks{Presented at
\href{https://web.cvent.com/event/1b7d7c3a-e9b4-409d-ae2b-284779cfe72f/summary}{Workshop
on the Science of Scientific-Software Development and Use}, Sponsored by U.S.\
Department of Energy, Office of Advanced Scientific Computing Research, Dec 13--15, 2021}

\maketitle

Our goal is to identify inhibitors and catalysts for productive long-term
scientific software development.  The inhibitors and catalysts could take the
form of processes, tools, techniques, environmental factors (like working
conditions) and software artifacts (such as user manuals, unit tests, design
documents and code). The effort (time) invested in catalysts will pay off in
the long-term, while inhibitors will take up resources, and can lower product
quality.

Developer surveys on inhibitors and catalysts will yield responses as
varied as the education and experiential backgrounds of the respondents. 
Although well-meaning, responses will predictably be biased.
For instance, developers may be guilty of the \emph{sunk cost fallacy},
promoting a technology they have invested considerable hours in learning,
even if the current costs outweigh the benefits. Likewise developers
may recommend against spending time on proper requirements, not as an
indication that requirements are not valuable, only that current practice
doesn't promote requirements~\cite{HeatonAndCarver2015}. Another perceived
inhibitor is time spent in meetings. For instance, the lack of visible
short-term benefits renders department retreats unpopular, even though
relationship building and strategic decision making may provide significant
future rewards. Evaluating the usefulness of meetings is difficult.
Rather than relying on preference and perception,
as these examples illustrate, \emph{we need to measure the long-term impact
of development choices to make wise ones.}

\section{Building Blocks}

A scientific approach requires a solid foundation.  The building blocks for
scientific discourse are: communicating concepts via an unambiguous language,
formulating hypotheses, planning data collection, and analyzing models and
theories.  To start with, we need to classify the software under discussion.
Likely dimensions include: general purpose scientific tools versus special
purpose physical models, scientific domain, open source versus commercial
software, project maturity, project size, and level of safety criticality.

We also need to be precise about our software quality goals. Qualities
such as reliability, sustainability, reproducibility and productivity need
precise definitions. Attempts have been made since the
1970s~\cite{McCallEtAl1977}, but the resulting definitions aren't usually
specific to scientific software (as shown by the confusion between precision
and accuracy is the ISO/IEC definitions~\cite{ISO9126}). Moreover, the
definitions often focus on measurability, where the first priority should be
conceptual clarity, analogous to the unmeasurable, but conceptually clear,
definition of forward error, which requires knowing the (usually unknown) true
answer.

For each relevant quality we recommend collecting as many distinct
definitions as possible.  Once collected, they can be
assessed against the following criteria (based on~\citet{IEEE1998}):
completeness, consistency, modifiability, traceability, unambiguity and
abstractness. The understanding gained from this systematic survey and analysis
can be used to either choose solid definitions, or propose new ones.
In all cases, the definitions should enable \emph{reasoning about quality}.

\section{Productivity}

Our definition of long-term productivity~\cite{SmithAndCarette2020arXiv}
provides an example of our vision, and meets our criteria.
We define productivity as:

$$P = O / I$$ 
$$ I = \int_{0}^{T} H(t)\ dt $$
$$ O = \int_{0}^{T} \sum_{c \in C} F(S_c(t), K_c(t))\ dt $$

\noindent where $P$ is productivity, $I$ is the inputs, $O$ is the outputs, $0$
is the time the project started, $T$ is the time \emph{in the future} where we
want to take stock, $H$ is the total number of hours available by all personnel,
$C$ represents different classes of users (external as well as internal), $S$ is
\emph{user satisfaction} and $K$ is \emph{effective knowledge}, and $F$ is a
weighing function that indicates ``value''.  Thus productivity is
measured in ``value per year.''  and is a mixture of external and internal
value produced. \emph{Value} should not be equated with money; measuring
the productivity of free software development is just as important as for
commercial software.

While the most straightforward use of such a formula is to measure productivity
of a team, it can also be used in ``what if'' scenarios to assist in planning
interventions, i.e.\ changes intended to improve productivity.

Measuring over too short a time-frame will assuredly give
warped results. This leads some to argue that productivity shouldn't even
be measured~\cite{Ko2019}.

\section{Measuring}

Proper science requires measurement.  We can only determine whether a given
intervention is a catalyst or inhibitor by measuring its impact.  Let us
examine in more details the consequences of our proposed definition.

First, the time integrals emphasize that productivity is something that happens
\emph{over time}. The most interesting kind of productivity is that of an
organization over the span of years. Measuring over too short a time frame is
one of the main sources of \emph{technical debt}~\cite{KruchtenEtAl2012} as it
devalues planning, team work, being strategic, etc.

Secondly, as Drucker~\cite{Drucker1999} reminds us, quality is at least as
important as quantity. Here we use a proxy for quality, namely
\emph{user satisfaction}. It is important to note that unreleased products
and unreleased features induce \textbf{no} user satisfaction. A broken
product might be even worse, and produce negative satisfaction.

The input $H$ is the number of hours worked by the team, including managers
and support staff, as appropriate. To optimize productivity, we want to
make $I$, and thus $H$, \emph{small}. This is the raw input being applied,
whether effective or not.

We use user satisfaction ($S$) as a proxy for effective quality. How to 
measure this is left for future study. It can be approximated by measures
such as numbers of users, number of citations, number
of forks of a repository, number of ``stars'', surveys of existing users,
number of mentions in the issue tracker, and usability experiments.

Probably the trickiest part is \emph{effective knowledge} ($K$). The idea
is that while source code embodies operational knowledge that has the
potential to directly lead to user satisfaction, a project usually also
generates a lot of \emph{tacit knowledge} about design, including the
rationale for various choices. This is the kind of knowledge that is lost
when employees leave, and is the most costly to build and replace.
In other words, human-reusable knowledge such as documentation factors in
here.  The best measure for knowledge is an area for future exploration.

\section{Artifacts Produced}

Software development typically produces many artifacts, such as
requirements, specifications, user manuals, unit
tests, system tests, usability tests, build scripts, API (Application
Programming Interface) documentation, READMEs, license documents, process
documents, and code. We regard all of these as containing
\emph{knowledge}, albeit encoded in different forms. Furthermore, it is
crucial to recognize that the knowledge of a single product is 
\emph{distributed} amongst those artifacts. In particular, the various
artifacts contain many copies of the same core knowledge --- by design.

To understand the importance of certain artifacts, it makes sense to look
at the productivity impact of their presence/absence. For example,
long-lived projects will inevitably encounter contributor turnover. How long
should it take for new contributors to be productive? How much training by
peer mentors will it take? Could some documentation be written that would
shorten this learning period and, just as importantly, reduce the time it
takes from experienced people? Of course, documentation that is out-of-date
could be even worse: a false sense of knowledge that results in even more
wasted work that needs repairing.

As we gain understanding on measures of value, we can use them to evaluate the
state of practice in different research software domains. We can estimate the
knowledge $K$ embedded in, and the user value $S$ derived from, existing
artifacts. In particular, we can compare these to the artifacts produced by
recommended processes from standard software engineering textbooks. For example,
we can test the hypothesis that knowledge duplication between code and
requirements, coupled with the fact that requirements get de-synchronized from
the code and the tenuous link to user value, is the likely reason for low
adoption of requirements in scientific software
development~\cite{HeatonAndCarver2015}.
 
Nevertheless, documentation remains useful, especially for the very long
term. Another means to judge the utility of documentation is to look at
assurance cases. An assurance case~\cite{RinehartEtAl2015} presents an
organized and explicit argument for correctness (or whatever other software
quality is deemed important) through a series of sub-arguments and evidence.
Assurance cases gives at least one measure of which documentation is relevant and
necessary.

\section{Production Methods}

One way to improve productivity is to waste less on non-productive or
counter-productive activities. That code is the most visible artifact that
contributes user-value, along with with testing (because quality is an
extremely important factor in user-value) explains the inordinate focus
on just those artifacts. Furthermore, the de-emphasis on documentation, even to
the extreme of some methodologies having none, can feel like productivity
improvements in the short term!  A better approach would be to capture
knowledge in ways that keeps it continuously synchronized between the
various artifacts where it appears.

One promising approach is to generate all artifacts from a single
knowledge base~\cite{SzymczakEtAl2016}.  This relies on a solid understanding
of the contents of all of the artifacts present in the software engineering
process. Our proof-of-concept shows that this is possible. As the artifacts
are now generated, knowledge duplication is not a problem. Even better, the
knowledge is synchronized-by-construction. Furthermore, it becomes easy to
tailor artifacts, documentation as well as code, to different classes of
``users''. 

\section{Concluding Remarks}

Our position is that decisions on processes, tools, techniques and software
artifacts should be driven by science, not by personal preference.  Decisions
should not be based on anecdotal evidence, gut instinct or the path of least
resistance.  Moreover, decisions should vary depending on the users and the
context.  In most cases of interest, this means that a longer term view should
be adopted.  We need to use a scientific approach based on unambiguous
definitions, empirical evidence, hypothesis testing and rigorous processes.

By developing an understanding of where input hours are spent, what most
contributes to user satisfaction, and how to leverage knowledge produced,
we can determine what has the greatest return on investment.  We will be able
to recommend software production processes that justify their value because the
long-term output benefits are high compared to the required input resources.  


\bibliographystyle{ACM-Reference-Format}
\bibliography{References}

\end{document}